\documentclass[aps,prl,twocolumn,showpacs,a4paper,superscriptaddress]{revtex4}

\usepackage{graphicx}
\usepackage{amsmath}
\begin{document}
\title{Quasiparticle Chirality in Epitaxial Graphene Probed at the Nanometer Scale}

\author{I.~Brihuega}
\affiliation{Max-Planck-Institut f\"{u}r Festk\"{o}rperforschung, Heisenbergstrasse 1, D-70569 Stuttgart, Germany\\}

\author{P.~Mallet}
\email[electronic address: ]{pierre.mallet@grenoble.cnrs.fr}
\affiliation{Institut N\'eel, CNRS-UJF, BP166, 38042 Grenoble, France\\}

\author{C.~Bena}
\affiliation{Institut de Physique Th\'{e}orique, CEA/Saclay, Orme des Merisiers, 91190 Gif-sur-Yvette, France\\}

\author{S.~Bose}
\affiliation{Max-Planck-Institut f\"{u}r Festk\"{o}rperforschung, Heisenbergstrasse 1, D-70569 Stuttgart, Germany\\}

\author{C.~Michaelis}
\affiliation{Max-Planck-Institut f\"{u}r Festk\"{o}rperforschung, Heisenbergstrasse 1, D-70569 Stuttgart, Germany\\}

\author{L.~Vitali}
\affiliation{Max-Planck-Institut f\"{u}r Festk\"{o}rperforschung, Heisenbergstrasse 1, D-70569 Stuttgart, Germany\\}

\author{F.~Varchon}
\affiliation{Institut N\'eel, CNRS-UJF, BP166, 38042 Grenoble, France\\}

\author{L.~Magaud}
\affiliation{Institut N\'eel, CNRS-UJF, BP166, 38042 Grenoble, France\\}

\author{K.~Kern}
\affiliation{Max-Planck-Institut f\"{u}r Festk\"{o}rperforschung, Heisenbergstrasse 1, D-70569 Stuttgart, Germany\\}
\affiliation{Institut de Physique des Nanostructures, Ecole Polytechnique F\'ed\'erale de Lausanne, CH-1015 Lausanne, Switzerland\\}

\author{J.~Y.~Veuillen}
\affiliation{Institut N\'eel, CNRS-UJF, BP166, 38042 Grenoble, France\\}

\date{published on 14th November 2008 in Physical Review Letters 101, 206802 (2008))}

%\date{\today}

\begin{abstract}
Graphene exhibits unconventional two-dimensional electronic properties resulting from the symmetry of its quasiparticles, which leads to the concepts of pseudospin and electronic chirality. Here we report that scanning tunneling microscopy can be used to probe these unique symmetry properties at the nanometer scale. They are reflected in the quantum interference pattern resulting from elastic scattering off impurities, and they can be directly read from its fast Fourier transform. Our data, complemented by theoretical calculations, demonstrate that the pseudospin and the electronic chirality in epitaxial graphene on SiC(0001) correspond to the ones predicted for ideal graphene.

\end{abstract}


%72.10.Fk Scattering by point defects, dislocations, surfaces, and other imperfections (including Kondo effect)

%73.20.-r Electron states at surfaces and interfaces

%68.37.Ef Scanning tunneling microscopy (including chemistry induced with STM)

\maketitle

	Graphene, one single layer of carbon atoms packed in a honeycomb structure, has two identical carbon atoms in each unit cell and thus two equivalent atom sublattices. This gives rise to an extra degree of freedom absent in conventional two-dimensional (2D) systems, and leads to exceptional electronic properties. The major consequence of the honeycomb structure is that the low energy quasiparticles of graphene are described by a Dirac-like Hamiltonian. The energy spectrum is thus linear, and consists of two Dirac cones centered at the opposite corners $K$ and $K'$ of the Brillouin zone. This is, however, insufficient to understand properly the specific electronic properties of graphene. Due to the presence of the two C atoms per unit cell, the quasiparticles have to be described by two-component wave functions, each component specifying the (complex) amplitude on each atomic sublattice. This results in an additional degree of freedom known as pseudospin. The projection of the pseudospin on the direction of the momentum defines the chirality of the quasiparticules; in graphene the direction of the quasiparticle pseudospin is parallel (positive chirality) or antiparallel (negative chirality) to their momentum \cite{Geim,Katsnelson}. Within one single Dirac cone, electrons (or equivalently holes) of opposite direction have opposite pseudospin, but the same chirality. In graphene bilayer with Bernal stacking, chiral Dirac fermions are also found, but they are massive and their chirality differs from the one found in graphene monolayer \cite{Geim,Katsnelson}.

In order to fully exploit graphene in future electronic devices it is of primary importance
to assess the chirality of the quasiparticles, since this feature is central to the transport properties.
It is responsible for the new ``chiral'' quantum Hall effects (QHE) measured in exfoliated monolayer and bilayer graphene \cite{Novoselov,Zhang,Novoselov2}. In epitaxial graphene multilayers on the SiC(000$\overline{1}$) face,
Shubnikov-de-Haas oscillations and weak antilocalization effects consistent with the electronic chirality of ``uncoupled'' graphene layers
\cite{Berger,Wu} have been reported, but surprisingly the QHE has not been yet observed \cite{de Heer}. Therefore, a direct
measure of the electronic chirality in a single epitaxial graphene layer is
essential. Here we show that the quasiparticles pseudospin - and the associated chirality -
can be probed at the nanometer scale with a low temperature scanning
tunneling microscope (STM). Our results on graphene monolayer on SiC(0001) are consistent with the electronic chirality of ideal graphene.

Epitaxial graphene was grown on a 6H-SiC(0001) substrate by thermal desorption of silicon at high temperatures, in ultra high vacuum (UHV) \cite{Forbeaux,Mallet}. Parameters were chosen to obtain a mixed surface with a similar coverage of monolayer and bilayer (with Bernal stacking) terraces, $\sim$ 50 nm wide. A carbon layer is present at the interface between graphene and the SiC. This layer is strongly bound to the substrate, and acts as a buffer layer with negligible interactions with the graphene layers \cite{Varchon}. Although not metallic, the carbon buffer layer has electronic states that give a significant contribution to the STM images \cite{Mallet,Rutter}. Our experiments were done using a home-made microscope operating at 4K in ultra-high vacuum. High bias STM images, as shown in Fig. 1(a), allow discriminating monolayer from bilayer graphene terraces due to the different interface states contribution \cite{Mallet}.

Two groups have reported angle-resolved photoemission spectroscopy (ARPES) measurements on epitaxial graphene layers on SiC(0001). They demonstrate that the graphene layers are electron doped. The measured difference between the Fermi energy ($E_{F}$) and the Dirac energy ($E_{D}$) is $\sim$ 0.45 eV for a monolayer and $\sim$ 0.30 eV for a bilayer \cite{Bostwick,Zhou,Ohta}. For graphene monolayer, a low energy linear dispersion has been found, a fingerprint of Dirac-like fermions \cite{Bostwick,Zhou}. However, different results are obtained by the two groups concerning the crossing of the bands: An energy gap of 0.25 eV is found at the Dirac point by only one of the groups \cite{Zhou}.
It has been proposed by the authors that the gap arises from a substrate-induced breaking of the sublattice symmetry of graphene.
ARPES, also shows that the Fermi surfaces (FSs) of monolayer and bilayer graphene on SiC(0001) are essentially identical \cite{Ohta,Bostwick2}.They consist of two tiny circular pockets, of radius $q_{F}$, surrounding the $K$ and $K'$ points [Fig. 1(b), six pockets are sketched to account for the symmetry of the Brillouin zone].

\begin{figure}
\includegraphics[width=6.3 cm,clip]{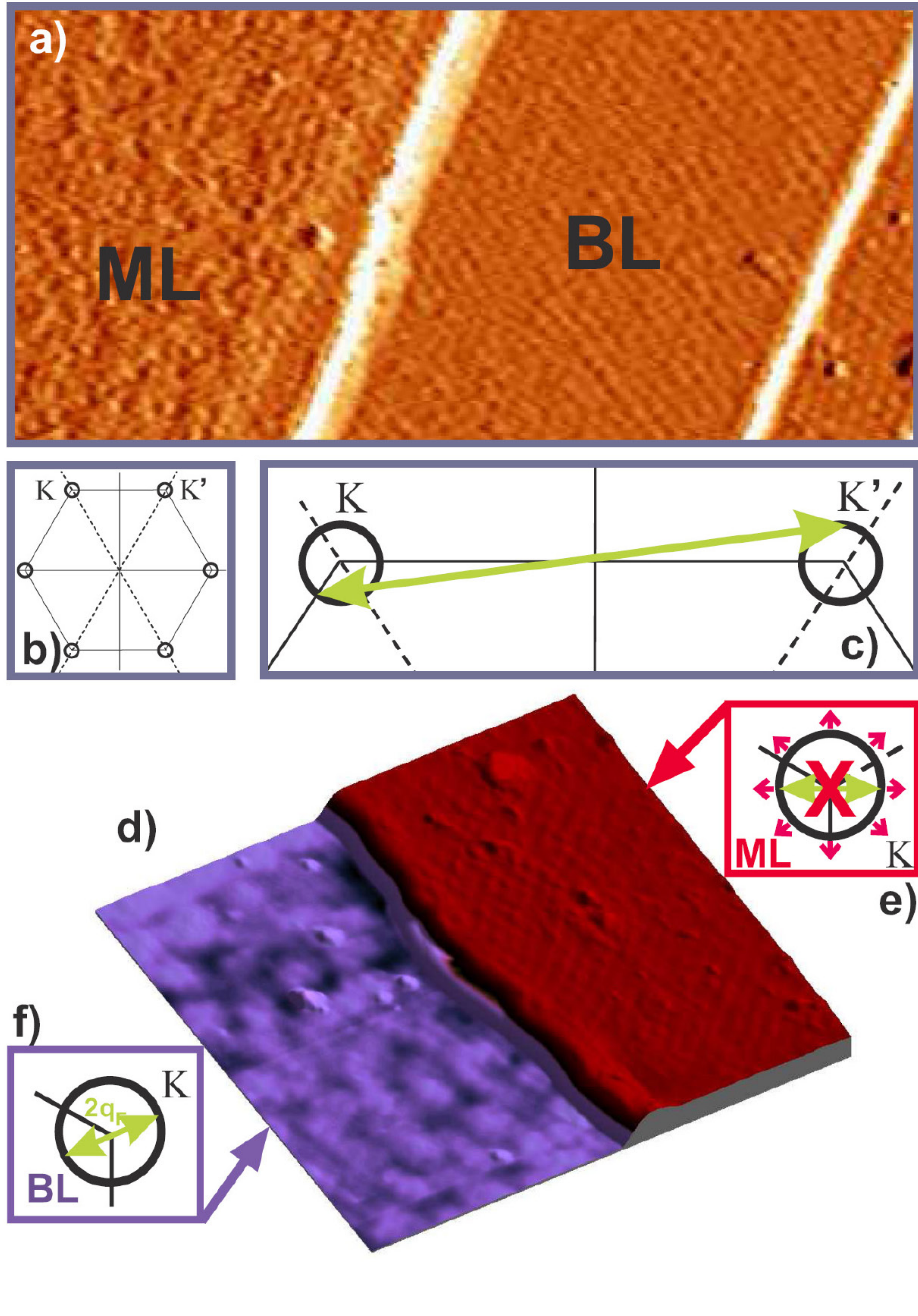}
\caption{(color online) (a) Constant current 90$\times$43 nm$^{2}$ STM image of epitaxial graphene on SiC(0001), with two adjacent
monolayer (ML) and bilayer (BL) graphene terraces. The spatial derivative of the image is
shown, to enhance the corrugation due to the interface states which is higher on ML terraces \cite{Mallet}.
Sample bias: +980 mV, tunneling current: 0.15 nA. (b) Schematic Fermi surface for electron-doped ML and BL
graphene. (c) Illustration of an intervalley scattering process. (d) 3D rendered 50$\times$50 nm$^{2}$
constant current image of two adjacent ML and BL terraces, taken at low sample bias (+ 1mV). A
long wavelength scattering pattern is found on the BL terrace (left), and not on the ML
terrace (right). Tunneling current: 0.2 nA. (e) Schematic of forbidden intravalley backscattering for ML graphene. Red
arrows sketch the pseudospin direction. (f) Schematic of intravalley backscattering for BL
graphene.}
\label{f.1}
\end{figure}

At low sample bias (a few mVs), the STM probes essentially the electronic states of the FS, and a constant current image reflects a map of the surface local density of states (LDOS) at $E_{F}$ \cite{Tersoff}. In a 2D system such as graphene, low bias images probe the quantum interferences (QIs) arising from elastic scattering of surface quasiparticles at static defects \cite{Crommie}. Such QIs are related to the Friedel oscillations in 2D metals. In a simple picture, any impurity scattering between two states $\overrightarrow{k}$ and $\overrightarrow{k}'$ of the FS should give rise to LDOS spatial modulations with wavevector $\overrightarrow{k}'- \overrightarrow{k}$, with a scattering probability depending on the topology of the FS. For a circular FS of radius $k_{F}$, scattering between states $\overrightarrow{k}_{F}$ and states $-\overrightarrow{k}_{F}$ (termed backscattering in the following) are the most efficient processes due to enhanced phase space, resulting in LDOS modulations with wavector $2k_{F}$ \cite{Crommie,Sprunger}.

For the FS sketched in Fig. 1(b), and depending on the impurity potentials present in the system, we may expect elastic electron scattering processes between the two non-equivalent Dirac cones at $K$ and $K'$ (i.e. intervalley scattering) as depicted in Fig. 1(c), together with elastic scattering processes between states located on the same Dirac cone (i.e. intravalley scattering). Figure 1(d) is a constant current STM image taken at +2mV that includes two adjacent graphene monolayer and bilayer terraces. A striking difference is measured in the LDOS patterns between the left and right terraces. A clear long-range modulation, of wavelength 5.2$\pm$0.3 nm, is found on the graphene bilayer terrace. This modulation has already been reported by Rutter et al. \cite{Rutter}, and is attributed to quantum interferences of wave vector 2$q_{F}$ due to intravalley backscattering between states $\overrightarrow{q }_{F}$ and states $-\overrightarrow{q }_{F}$ within one single pocket [Fig. 1(f)].  Surprisingly, we do not find such LDOS modulation on the monolayer, which is puzzling if we keep in mind that the two systems have almost the same FS.

The lack of LDOS modulation of wave vector 2$q_{F}$  for the monolayer implies that QIs associated with
intravalley backscattering are virtually absent. It has been reported that wave function symmetries may lead to selection rules that can prevent QIs, for instance in systems with strong spin-orbit coupling where the electronic spin governs the elastic scattering processes \cite{Petersen,Pascual}. In the present case, the absence of QIs in graphene monolayer is associated to the pseudospin instead of real spin.
Intuitively, this is because the pseudospin of the backscattered wave is
opposite to that of the incident wave, so that their overlap should be zero [see Fig. 1(e)].
For a scattering potential with range larger than
the C-C distance (which leads to scattering without pseudospin-flip),
it has been shown that intravalley backscattering
processes and the associated QIs  are strictly forbidden in
monolayer graphene \cite{Ando,Katsnelson2}.
On the contrary, the electronic chirality does not prevent
intravalley backscattering in bilayer graphene [Fig. 1(f)] \cite{Katsnelson3}.

For both monolayer and bilayer epitaxial graphene on SiC(0001), previous STM
experiments have reported that atomic-size impurities give rise to short-wavelength
modulations of the LDOS, associated with intervalley scattering \cite{Mallet,Rutter}. According to
Ando et al. \cite{Ando}, such impurities, acting as short-range potentials, should also
allow intravalley backscattering, which raises the question of possible QIs of wave vector
2$q_{F}$. Recent theoretical studies show that for a general impurity potential, LDOS
modulations with wave vector $2q_{F}$ should exist for graphene monolayer, but they
should be strongly attenuated \cite{Bena,Mariani,Cheianov}, with a $1/r^{2}$ decay law instead of
the $1/r$ law found in conventional 2D systems. On the other hand, a $1/r$ decay law is
obtained for the QIs due to intravalley backscattering in a graphene bilayer, and also for the
QIs due to intervalley scattering for both monolayer and bilayer graphene. In reference \cite{Bena} one of us showed that the $1/r^{2}$ decay law should have a clear impact on the 2D Fourier transform (FT) of the LDOS modulations: while on
bilayer graphene the efficiency of intravalley backscattering should give rise to a central
ring of radius 2$q_{F}$ in the FT map similar to the one observed for conventional 2D systems,
the strong attenuation of intravalley backscattering on monolayer graphene should generate
a FT image with no central ring \cite{Bena}.

  \begin{figure}
\includegraphics[width=8.4 cm,clip]{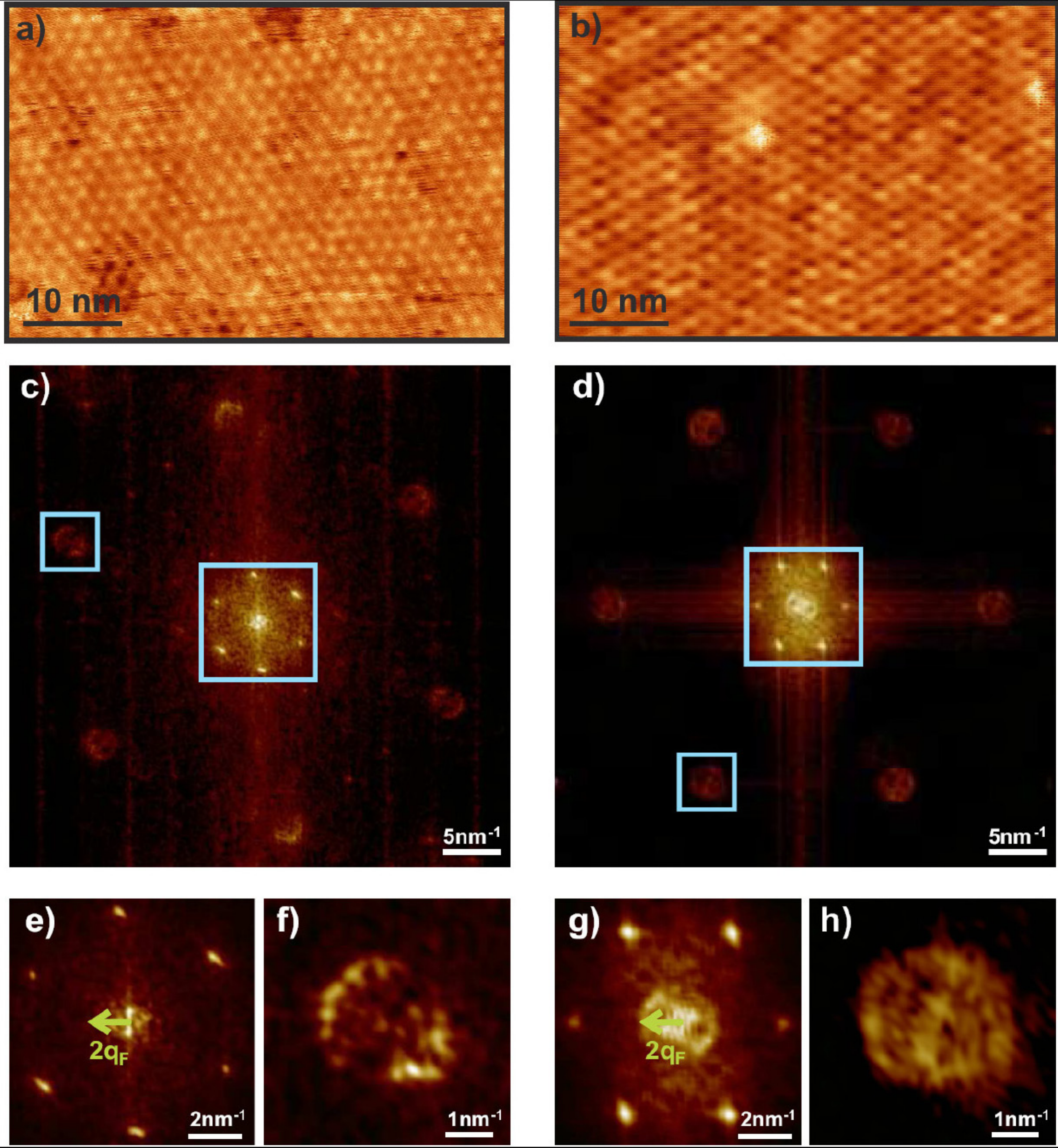}
\caption{(color online) (a),(b) Low-bias STM images of 50 nm wide monolayer (a) and bilayer (b)
terraces. Sample bias and tunneling current are respectively +2 mV and 0.4 nA for (a), +4 mV and 0.13 nA for (b). (c),(d) Two-dimensional fast
Fourier transform (FFT) maps of the STM images (a) and (b). (e) Central
region of (c), showing no intravalley-backscattering related ring (the green arrow points out
the position where such a ring should appear). (f) One of the outer pockets of
(c). (g) Central region of (d), showing a clear ring-like feature of radius
2$q_{F}$ related to intravalley-backscattering. (h) One of the outer pockets of (d).
Outer pockets shown in (f) and (h) are centered at the $K$ (or $K$') point and result from intervalley scattering.}
\label{f.2}
\end{figure}

In the following, we present high resolution fast Fourier transform (FFT) of STM
images recorded at low sample bias on 50 nm wide graphene terraces. We first show the real-space
images, for graphene monolayer [Fig. 2(a)] and
graphene bilayer [Fig. 2(b)]. Note that both images exhibit a triangular pattern of periodicity $\sim$ 1.9 nm
which is related to the interface reconstruction \cite{Mallet,Rutter}, and which appears as a sextuplet of bright
spots in the corresponding FFT images [Figs. 2(c) and 2(d)].

The central region of the FFT in Figs. 2(c) and 2(d) is related to intravalley
scattering: a clear ring-like feature of average radius 1.2 nm$^{-1}$ is found for the bilayer [Figs.
2(d) and 2(g)]. This radius value is in agreement with the value given in ref. \cite{Rutter}, and with
the 2$q_{F}$ value derived from ARPES \cite{Bostwick,Zhou}. On the monolayer terrace, no central ring is
found [Figs. 2(c) and 2(e)], despite the unprecedented momentum resolution obtained here.
This has been checked on many different monolayer terraces (see also the EPAPS document \cite{EPAPS}), and we conclude
that the result is robust and systematic. As discussed above, pseudospin
and electronic chirality are the key parameters for the absence (presence) of the central ring
in the FFT image of monolayer (bilayer) graphene.

In the high frequency regions of the FFTs of Figs. 2(c) and 2(d), we find six outer pockets
with ring-like shapes centered at $K$($K'$) points. They result from intervalley
scattering, associated to real space LDOS modulations with a ($\sqrt{3}\times\sqrt{3})R30^{\circ}$ periodicity
with respect to graphene \cite{Mallet,Rutter}. As shown in Figs. 2(f) and 2(h), the intensity of
the high frequency rings in the FFT is not isotropic. The anisotropy is much more
pronounced for graphene monolayer, for which a splitting of each ring into two arcs is found
[Fig. 2(f))]. This splitting is discussed in more detail in the EPAPS document \cite{EPAPS}.

The experimental results are compared with T-matrix calculation made along the lines
of Ref. \cite{Bena} (see also the EPAPS document \cite{EPAPS}). Figure 3(a) is the FT
of a calculated LDOS map at $E_{F}$ for monolayer graphene
with $E_{D}$ = $E_{F}$-0.45 eV, in the presence of a single localized impurity. It is clear
that no central 2$q_{F}$ ring is generated in this case. This is
the signature of LDOS modulations resulting from inefficient intravalley backscattering,
and it is the most direct fingerprint of graphene's pseudospin and chirality \cite{Bena}.
This is in very good agreement with the experimental absence of the central ring [Fig. 2(e)].
Also, the high frequency split rings,
resulting from intervalley scattering, are nicely reproduced by the
theoretical calculation. The splitting of the rings is another
consequence of the quasiparticle chirality \cite{EPAPS}.

\begin{figure}
\includegraphics[width=8.5cm,clip]{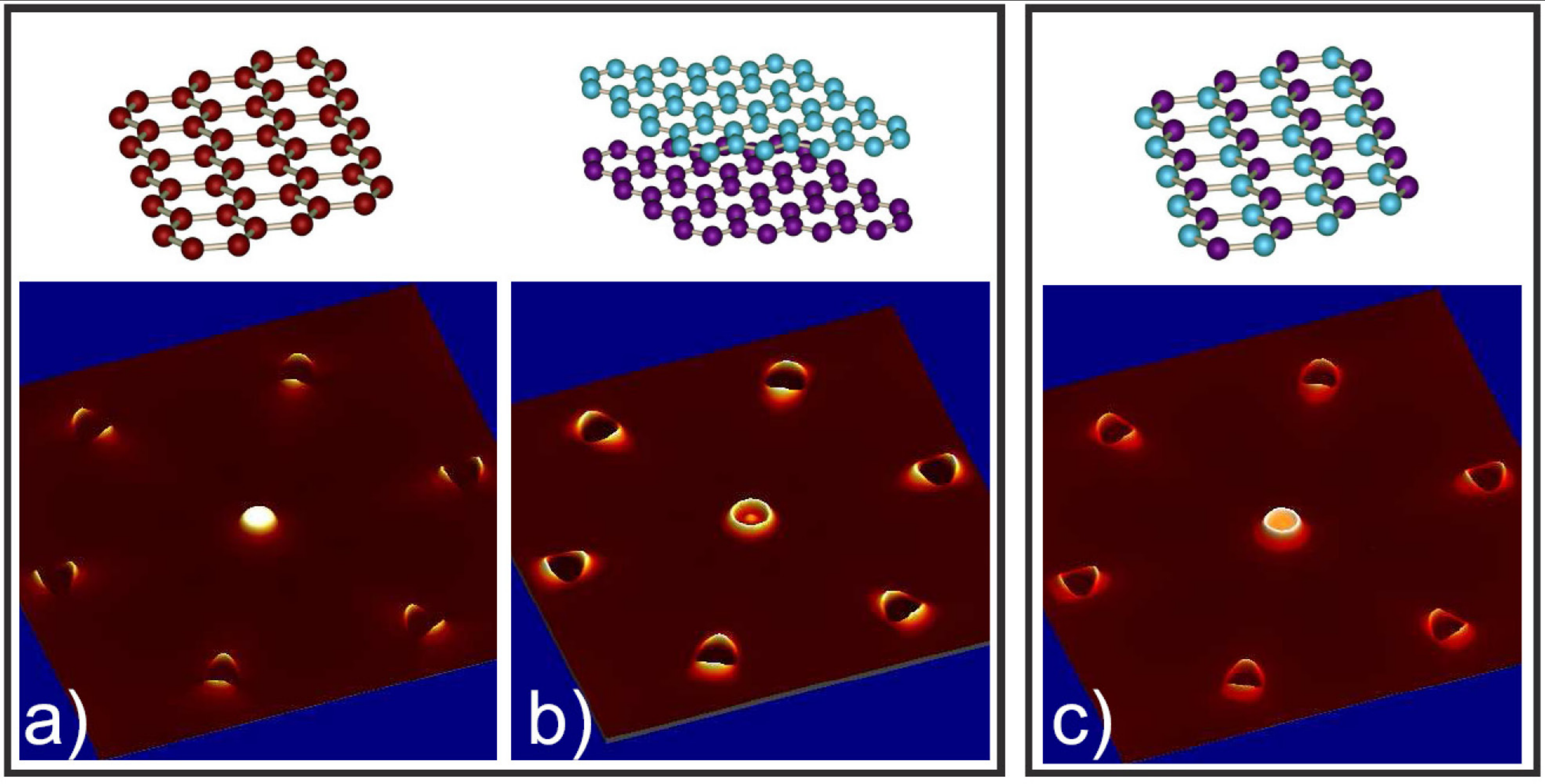}
\caption{(color online) (a-c) Theoretical calculation of the Fourier Transform (FT) of the LDOS at
$E_{F}$ for different graphene layers with a single delta-function impurity.
(a) 3D rendered FT map for graphene monolayer with a Dirac energy $E_{D}$ = $E_{F}-$0.45 eV.
 The
calculation predicts no central ring of radius $2q_{F}$,
showing the unefficiency of intravalley backscattering.
 (b) Same representation as (a) for a bilayer, showing a clear
central $2q_{F}$ ring. $E_{D}$ = $E_{F}-$0.3 eV. An asymmetry of $\sim$ 0.1eV between the two layers has
been included, to account for the different electron density reported by ARPES (12,13). (c)
Same representation as (a) for an asymmetric monolayer, with $E_{D}$ = $E_{F}-$0.45eV. A site
energy difference of 0.25 eV is introduced between the carbon atoms of each unit cell, and
the impurity is placed on a carbon atom with high site energy. The symmetry breaking
between the two carbon sublattices restores the efficiency of intravalley backscattering as
shown by the reappearance of a central ring in the FT map.}
\label{f.2}
\end{figure}

Figure 3(b) is an equivalent map as Fig. 3(a) but for a graphene bilayer, with $E_{D}$ = $E_{F}$-0.3 eV.
 From ARPES it is known that the two layers have a different electron density,
giving rise to a gap at $E_{D}$ of about 0.1 eV \cite{Ohta}. The asymmetry between the two layers has
been included in the present calculation. Note the appearance of the inner ring of radius
2$q_{F}$, as well as the presence of the outer rings which have not obvious splitting (a faint
intensity remains perpendicularly to the $\Gamma K$ and $\Gamma K'$ directions). The agreement with the
experimental FFT is excellent.

Finally, we propose the use of STM to check any possible breaking of graphene's
sublattice symmetry. To this end, we have calculated the FT of the LDOS map for a
monolayer with the same parameters as in Fig. 3(a), but including an arbitrary sublattice asymmetry of 0.25 eV,
which is the gap value measured by Zhou et al.
\cite{Zhou}. The calculation shows a clear inner ring in the FT, corresponding to efficient QIs due
to intravalley backscattering \cite{EPAPS}. This dramatically illustrates the effect of the asymmetry
 on the FT STM data since this ring is absent in the symmetrical case [Fig. 3(a)].
The data of Fig. 3(c) are for a defect on one sublattice,
 but we found that the shape of the ring depends on the location of the scattering potential \cite{EPAPS}.
  Due to the large size of the images it is likely that a distribution of defects is present in the experimental data. Calculations
  indicate that a large sublattice asymmetry (leading to a gap of 0.6 eV) should be readily detected in FT-STM images, but that for a moderate asymmetry (as for Fig. 3(c)) the intensity of the ring could depend on the defect configuration. Since we have taken images in many different surface areas without detecting any ring, we believe that our data are rather consistent with a symmetrical monolayer, although a small asymmetry can not be totally ruled out.

In summary, by analyzing the Fourier transform of low bias STM images, we have
demonstrated that the pseudospin and the chirality of epitaxial graphene quasiparticles can be
probed directly at the nanometer scale. We show that graphene monolayer on SiC(0001)
exhibits the chirality of an ideal graphene sheet. Together with the previous
evidence by ARPES of the low energy linear spectrum, a complete assessment of the Dirac nature of the quasiparticles in this
system is obtained.

We thank D. Mayou for helpful discussions, F. Hiebel for participating to sample
preparation, R. Cox for careful reading of the manuscript and the French ANR (programme
Blanc) for partial financial support of this work. C. B. and I. B. were supported by a Marie
Curie action under the Sixth Framework Programme.

%\newpage
%\begin{figure}
%\includegraphics[angle=-90, width=6.0cm,clip]{cell.ps}
%\caption{interface geometry. a- side view, b- top view of the
%$\sqrt{3}\times \sqrt{3}R30$ cell in the case of a Si-terminated
%SiC face. The lonely atom is missing in the C-deficient geometry}
%\label{f.1}
%\end{figure}

%\begin{figure}
%\includegraphics[width=8.0cm,clip]{Reflectivity_fig.eps}
%\caption{X-ray $(00l)$ reflectivity data from ~9 graphene layers
%grown on the 4H-SiC$(000\bar{1})$ surface. Bulk and graphite Bragg
%peaks are labelled. Solid line is a fit as described in the text.}
%\label{f.2}
%\end{figure}

%\begin{figure}
%\includegraphics[angle=-90, width=12.0cm,clip]{bands.ps}
%\caption{dispersion curves for one (a,b), two (c,d) and three
%(e,f) graphene layers on bulk truncated SiC. a,c,e corespond to Si
%terminated face; b,d,f to C terminated face. Inset in c shows a zoom of the anticrossing in the vicinity of $E_F$.} %\label{f.3}
%\end{figure}

%\begin{figure}
%\includegraphics[angle=-90,width=5.0cm,clip]{Cdef.ps}
%\caption{dispersion curves for 2 C layers on top of the
%C-deficient surface} \label{f.4}
%\end{figure}
%\begin{figure}
%\includegraphics[width=8.0cm,clip]{isocharge.eps}
%\caption{isocharge density contours along z axis for three C
%layers on C- (a) and Si-(b) terminated surface} \label{f.5}
%\end{figure}

\end{document}